\documentclass[aps,prb,superscriptaddress,showpacs,twocolumn]{revtex4}

\usepackage{graphicx}
\usepackage{bm}
\usepackage{amsmath}

\newcommand{\beq}{\begin{equation}}
\newcommand{\eeq}{\end{equation}}
\newcommand{\beqa}{\begin{eqnarray}}
\newcommand{\eeqa}{\end{eqnarray}}

\newcommand{\ket}[1]{\left| #1 \right\rangle}
\newcommand{\bra}[1]{\left\langle #1 \right|}

\begin{document}
\title{\bf Photon statistics from a resonantly driven quantum dot}
\author{Ahsan Nazir} \email{ahsan.nazir@ucl.ac.uk}
\affiliation{Centre for Quantum Dynamics and Centre for Quantum Computer Technology, Griffith University, Brisbane, Queensland 4111,
Australia}
\affiliation{Department of Physics and Astronomy, University College London, London WC1E 6BT, U. K.}

\date{\today}

\begin{abstract}
Photon correlations in the emission of a resonantly 
driven quantum dot are investigated, accounting for the influence of the solid-state phonon environment. An analytical expression is derived for the second-order fluorescence intensity correlation function, from which regimes of correlated and uncorrelated photon emission are predicted as the driving field is varied. Experiments to investigate this effect would provide valuable insight into quantum dot carrier-phonon dynamics and are feasible with current technology.
\end{abstract}


\maketitle

There has been a great deal of progress over the past few years in refining fabrication and characterisation techniques for samples of small semiconductor quantum dots (QDs).~\cite{Krenner2005,Shields2007} This has lead to the ability to control and manipulate, with ever increasing precision, the quantum states of charge carriers confined within single QDs,~\cite{Bonadeo98} and QD arrays.~\cite{Unold05} Such systems are prototype solid-state quantum processors, ideal for performing studies of coherent control techniques, decoherence, and quantum measurements that will be key in assessing the prospects for larger scale solid-state quantum information processing (QIP). QDs are also promising candidates for efficient single photon sources,~\cite{Michler00,Shields2007} opening up potential applications in other paradigms such as linear optical~\cite{Kok07} and hybrid light-matter QIP.~\cite{BarrettKok2005} 

There is therefore considerable interest in employing QDs both as qubits themselves and as elements in a larger optical setup. For either application an understanding of the coherence properties of their optical transitions is key. In the former case to ensure high fidelity qubit operations, in the latter case to evaluate the degree of photon indistinguishability in the QD emission. Hence, an experiment that establishes both the nonclassical nature of the QD fluorescence, while at the same time exploring the potential for coherent control of the QD excitonic states themselves, is desirable. In fact, a first experiment along exactly these lines has recently been reported by Muller {\it et al.},~\cite{Muller07} who measured the second-order fluorescence intensity correlation function, $g^{(2)}(\tau)$, for a coherently-driven InGaAs QD, observing damped oscillations in the signal.      

Here, I shall show that not only does such a measurement provide a clear signature of a solid-state two-level quantum emitter, but that it is also a sensitive probe of the coherence properties of the QD excitonic ground-state. In contrast to the excitonic occupation dynamics usually investigated,~\cite{Machnikowski04,Vagov07,Forstner03,Mogilevtsev08} in the following I shall derive an analytical expression for the fluorescence intensity correlation function of a driven QD. I shall use it to show how the interplay between laser-induced coherent exciton evolution, spontaneous photon emission, and dephasing due to exciton-phonon interactions~\cite{Borri05} manifests itself in the  emitted photon statistics. In particular, it is expected that the phonon-induced dephasing rate should display a non-monotonic driving strength dependence.~\cite{Machnikowski04,Vagov07,Forstner03,Mogilevtsev08} This can be understood from a general resonance argument;~\cite{Machnikowski04,Vagov07} the induced damping is dominated by phonons that are most strongly coupled to the QD carriers, which corresponds to those that have wavelengths comparable to the QD size $d$ (or frequencies $\sim u/d$, where $u$ is the sound velocity~\cite{Vagov07}). These modes are most effectively driven when the carrier dynamics is resonant with them, which in turn implies that the Rabi frequency (QD-driving coupling strength) is resonant with them too. Hence, as the Rabi frequency is increased towards a value of $u/d$ the carrier-phonon coupling increases as well, leading to a larger dephasing rate, only for it to decrease again as the Rabi frequency is increased further, moving the carrier dynamics out of resonance with the dominant phonon modes. I shall demonstrate below that these dephasing-rate variations give rise to a distinctive feature in the second-order correlation function, namely the appearance, disappearance, and subsequent reappearance of long-lived oscillations in $g^{(2)}(\tau)$ as the driving field is changed.

We consider a single QD modelled by a two-level system with ground-state $|0\rangle$  (the semiconductor ground-state) and excited state $|X\rangle$ (a single exciton), separated by an energy $\omega_0$ ($\hbar=1$). The dot is addressed by an external classical laser field of frequency $\omega_l$ and dot-laser coupling given in the dipole approximation by $\Omega$, with $\Omega\ll\omega_0$. Both the phonon environment and the radiation field are modelled by harmonic oscillator baths of frequencies $\omega_{\bf q}$ and $\theta_{\bf k}$, respectively, with ${\bf q}$ and ${\bf k}$ the corresponding wavevectors. The carrier-phonon interaction acts only on the dot excited state $\ket{X}$ and is taken to be of the usual spin-boson form.~\cite{Leggett87,Krummheuer02,Pazy02,Machnikowski04,Forstner03,Vagov07} Within a frame rotating at frequency $\omega_l$ this leads to the Hamiltonian:~\cite{Rodriguez02}
$H=\nu\ket{X}\bra{X}+\frac{\Omega}{2}(\ket{0}\bra{X}+\ket{X}\bra{0})+\sum_{\bf q}\omega_{\bf q}b_{\bf q}^{\dagger}b_{\bf q}+\ket{X}\bra{X}\sum_{\bf q}(g_{\bf q}b_{\bf q}^{\dagger}+g_{\bf q}^*b_{\bf q})+\sum_{\bf k}\theta_{\bf k}a_{\bf k}^{\dagger}a_{\bf k}+\sum_{\bf k}(f_{\bf k}\sigma_-a_{\bf k}^{\dagger}+f_{\bf k}^*\sigma_+a_{\bf k})$,
with rotating-wave approximations on both the driving and system-photon coupling terms. Here, $\nu=\omega_0-\omega_l$ is the detuning of the laser from the dot transition, $\sigma_-=\ket{0}\bra{X}$, $\sigma_+=\ket{X}\bra{0}$, $g_{\bf q}$ ($f_{\bf k}$) defines the exciton-phonon (exciton-photon) coupling, $b_{\bf q}$ ($a_{\bf k}$) are the phonon (photon) annihilation operators, and an irrelevant term has been dropped.

On taking a trace over the phonon modes, which are assumed to be in thermal equilibrium, the resulting spectral density $J(\omega)=\sum_{\bf q}|g_{\bf q}|^2\delta(\omega-\omega_{\bf q})$ completely describes the interaction between the QD and phonons.~\cite{Leggett87} For arsenide QDs, the coupling of the confined exciton to acoustic phonons by means of the deformation potential tends to dominate the dephasing dynamics, over the piezoelectric interaction or coupling to optical phonons.~\cite{Borri05,Krummheuer02,Pazy02} In this case, the coupling constant is given by $g_{\bf q}=qD({\bf q})/\sqrt{2\lambda\omega_{\bf q}V}$,~\cite{Mahan00,Calarco03} where $\lambda$ is the sample density and $V$ the unit cell volume. The form factor is $D({\bf q})=\int d{\bf r}\left(D_h|\psi_h({\bf r})|^2-D_e|\psi_e({\bf r})|^2\right)e^{-i{\bf q}\cdot{\bf r}}$, with $\psi_h$ and $\psi_e$ being the confined hole and electron ground-states, respectively, while $D_h$ and $D_e$ are the corresponding bulk deformation potential constants. For clarity, I take a spherically symmetric harmonic confinement potential for the QD, giving $\psi_{e(h)}=(d_{e(h)}\sqrt{\pi})^{-3/2}\exp{(-r^2/2d_{e(h)}^2)}$, where $d_{e(h)}$ is the electron (hole) ground-state localization length. For $d_e=d_h=d$ we obtain $J(\omega)=\omega^3(D_h-D_e)^2e^{-\frac{\omega^2d^2}{2u^2}}/(4\pi^2\lambda u^5)=\alpha\omega^3e^{-(\omega/\omega_c)^2}$, where a linear dispersion $\omega_{\bf q}=uq$ has been assumed. The spectral density is therefore super-ohmic with a natural high frequency cut-off at $\omega\sim\omega_c= \sqrt{2}u/d$ due to the finite QD size.

Under resonant continuous-wave excitation the system reaches a quasi steady-state (ss),~\cite{Muller07} after which time 
the normalised second-order fluorescence intensity correlation function of the QD, $g^{(2)}(\tau)=\langle I(t)I(t+\tau)\rangle/\langle I(t)\rangle^2$, may be written~\cite{Carmichael93}
\begin{equation}\label{g2definition}
g^{(2)}(\tau)=\langle\sigma_+\sigma_-\rangle_{\rm ss}^{-1}\left(1+\langle\sigma_z(\tau)\rangle_{\rho(0)=\ket{0}\bra{0}}\right).
\end{equation} 
This is proportional to the probability of detecting two emitted photons separated by a time $\tau$, and normalised by the form for independent detection. Here, and in the following, $\sigma_i$ (for $i=x,y,z$) are the usual Pauli matrices in the basis $\{|0\rangle,|X\rangle\}$.
Note that the evolution of $\sigma_z(\tau)$ must be obtained conditional on the system being prepared in its ground-state at $t=0$ ($\rho(0)=\ket{0}\bra{0}$), i.e. detection of the first photon initializes the system state. The continuous-wave excitation condition ensures that there is no effect of any laser-pulse temporal profile in subsequent measurements.

During their evolution the QD states are coupled both to a bath of phonons and the radiation field. Utilising the time-convolutionless projection operator technique~\cite{breuer02} a master equation for the reduced system density operator may be derived from the Hamiltonian and the Liouville-von Neumann equation, here within the Born approximation and up to second order in $|g_{\bf q}|$ and $|f_{\bf k}|$. Typically, the memory time within the radiation field is extremely short, $\tau_m\sim(1/\omega_0)\sim1$~fs for $\omega_0$ of around $1$~eV.~\cite{Hohenester04} Since $\Omega\ll\omega_0$, we are then justified in treating spontaneous emission within the Markov approximation as being governed by a constant rate $\gamma$.~\cite{Carmichael93,breuer02,Hohenester04}
The master equation thus has the following form:
\begin{align}\label{masterequation}
\dot{\rho}&=-i(H_{\rm eff}\rho-\rho H_{\rm eff}^{\dagger})-\frac{1}{4}\big(D(t)[\sigma_z,[\sigma_z,\rho]]-\xi_1(t)\sigma_x\rho\sigma_z\nonumber\\
&\quad-\xi_1^*(t)\sigma_z\rho\sigma_x-\xi_2(t)\sigma_y\rho\sigma_z-\xi_2^*(t)\sigma_z\rho\sigma_y\big)+\gamma\sigma_-\rho\sigma_+,
\end{align}
where $H_{\rm eff}=\left[(\Omega/2)+(\xi_2(t)/4)\right]\sigma_x-(\xi_1(t)/4)\sigma_y-(\nu'(t)/2)\sigma_z-(i\gamma/2)\sigma_+\sigma_-$ is non-Hermitian. 
Defining the bath correlation function $C(t)=\int_0^{\infty}d\omega J(\omega)\left(\cos{(\omega t)}\coth{(\beta\omega/2)}-i\sin{(\omega t)}\right)$, the following time-dependent rates are obtained:
\begin{subequations}\label{ratestimedep}
\begin{align}
D(t)&=\frac{1}{E^2}\int_0^tdt'{\rm Re}\{C(t')\}(\nu^2+\Omega^2\cos{(Et'))},\label{rateD}\\
\xi_1(t)&=\frac{\Omega\nu}{E^2}\int_0^tdt'C(t')(1-\cos{(Et'))},\label{ratexi1}\\
\xi_2(t)&=\frac{\Omega}{E}\int_0^tdt'C(t')\sin{(Et')}\label{ratexi2},
\end{align} 
\end{subequations}
where $E^2=\nu^2+\Omega^2$. The environment-shifted detuning entering $H_{\rm eff}$ is $\nu'(t)=\nu+\int_0^tdt'{\rm Im}\{C(t')\}$, and $\beta=1/k_{B}T $, where $k_B$ is Boltzmann's constant and $T$ the temperature. The effect of the phonon bath is therefore to induce frequency shifts, in $\nu'(t)$ and through the real parts of $\xi_{1(2)}(t)$ (arising from vacuum and thermal fluctuations), as well as irreversible terms through the dephasing $D(t)$ and the imaginary parts of $\xi_{1(2)}(t)$ (damping terms not present for $\Omega=0$).~\cite{Paz01}

\begin{figure}[!t]
\centering
\includegraphics[width=2.5in,height=1.4in]{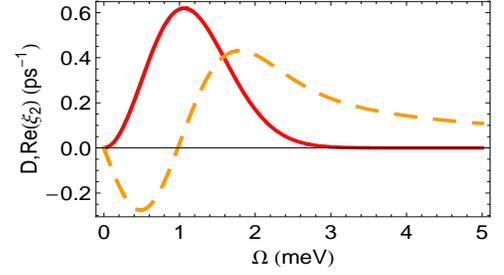}
\caption{(Color online) $D$ (red, solid line) and ${\rm Re}(\xi_2)$ (yellow, dashed line) as a function of driving strength $\Omega$. Parameters: $D_e=-14.6$~eV, $D_h=-4.8$~eV, $d=4.5$~nm, $u=5110$~${\rm ms}^{-1}$, $\lambda=5370$~kg${{\rm m}^{-3}}$, and $T=50$~K, giving $\alpha=0.032$~$\rm ps^2$.}
\label{figrates}
\end{figure}

As written above, Eq.~(\ref{masterequation}) is non-Markovian in the phonon terms, but local in time. Of primary interest here is the long-time behaviour of the second-order correlation function, measured under continuous excitation conditions. In this case, transients in the dynamics are ignored and the upper limits of integration in Eqs.~(\ref{rateD})~-~(\ref{ratexi2}) taken to infinity (a Markov approximation~\cite{breuer02}). The rates then become time-independent, the key point being that they are still functions of the laser detuning $\nu$ and dot-laser coupling $\Omega$.
In fact, only $D(\infty)=D$ and ${\rm Re}(\xi_2(\infty))={\rm Re}(\xi_2)$ will enter the final expression for $g^{(2)}(\tau)$, here given by $D=(\pi\Omega^2/2E^2)J(E)\coth{(\beta E/2)}=(\alpha\pi\Omega^2 E/2)e^{-(E/\omega_c)^2}\coth{(\beta E/2)}$, and for elevated temperatures
($k_BT>\omega_c$), ${\rm Re(\xi_2)}=\frac{\sqrt{\pi}\alpha\Omega}{24\beta}[2\sqrt{\pi}Ee^{-\left(\frac{E}{\omega_c}\right)^2}(12+E^2\beta^2){\rm erfi}(E/\omega_c)-\omega_c(24+(2E^2+\omega_c^2)\beta^2)]$, where ${\rm erfi}(z)$ is the imaginary error function. In addition, an expression for the bath-shifted detuning is needed; $\nu'=\nu-\int_0^{\infty}d\omega J(\omega)/\omega=\nu-\alpha\sqrt{\pi}\omega_c^3/4$.

Both $D$ and ${\rm Re}(\xi_2)$ are plotted against $\Omega$ in Fig.~\ref{figrates} for the resonance condition $\nu'=0$, i.e. $\nu=\alpha\sqrt{\pi}\omega_c^3/4$. In particular, the dephasing rate $D$ grows quickly as the driving strength is increased from zero ($D\propto\Omega^2$ at small $\Omega$) until the cut-off begins to dominate. Consequently, any excitonic oscillations induced by the coherent excitation become strongly damped as the driving frequency approaches the cut-off, 
as anticipated by the earlier resonance arguments. For example, when $\Omega=\omega_c\sim 1$~meV, we have $1/\Omega=1.5$~${\rm ps}^{-1}$, compared to $D=0.6$~${\rm ps}^{-1}$. This damping weakens however, and eventually becomes negligible, as $\Omega$ increases further, beyond $\omega_c$.

The resulting dynamics is clearly illustrated from the matrix form of Eq.~(\ref{masterequation}), $\langle\dot{\bf s}\rangle=M\langle{\bf s}\rangle-{\bf k}$, in terms of the Bloch vector ${\bf s}=(\sigma_-,\sigma_+,\sigma_z)^T$, with
\begin{equation}\label{blochvectorform}
M=
\left(
\begin{array}{ccc}
  -(D+\gamma/2) & 0  & i\Omega'-{\rm Re}(\xi_1)  \\
 0 & -(D+\gamma/2)  &  -i\Omega'-{\rm Re}(\xi_1) \\
 i\Omega/2 & -i\Omega/2  & -\gamma  
\end{array}
\right)\nonumber
\end{equation}
and ${\bf k}=(i{\rm Im}(\xi_1)+{\rm Im}(\xi_2),{\rm Im}(\xi_2)-i{\rm Im}(\xi_1),-\gamma)^T$. Here, $\Omega'=\Omega+{\rm Re}(\xi_2)$ and $\nu'=0$.
The dynamic evolution of $\langle\sigma_z(t)\rangle$ may now be found simply by solving for $\langle{\bf s}\rangle$, while setting $\langle\dot{{\bf s}}\rangle=0$ gives the required steady-state solution for $\langle\sigma_+\sigma_-\rangle$. From Eq.~(\ref{g2definition}) we then find an analytical expression for the normalised second-order correlation function
\begin{equation}\label{g2eqn}
g^{(2)}(\tau)=1-e^{-(2D+3\gamma)\tau/4}\left(\cos{\Delta \tau}+\frac{2D+3\gamma}{4\Delta}\sin{\Delta \tau}\right),\nonumber
\end{equation}
describing the detection statistics of photons emitted from the resonantly driven, dephasing QD,
with $\Delta=\sqrt{\Omega(\Omega+{\rm Re}(\xi_2))-(2D-\gamma)^2/16}$ now the effective oscillation frequency.
The correlation function is always zero at $\tau=0$, displaying the expected antibunching dip of a nonclassical emitter.~\cite{Carmichael93} As $\tau$ increases beyond the characteristic dephasing time (i.e. beyond ${\rm min}(1/D,1/\gamma)$), $g^{(2)}(\tau)\rightarrow 1$ indicating independent photon detection due to dephasing of the optical transition. At intermediate times, the correlation function describes an exponentially damped oscillation provided that $\Delta$ is real. For a typical QD this is usually the case beyond weak driving fields, here satisfied when $\Omega>1$~$\mu{\rm eV}$ for the parameters of Fig.~\ref{figrates}. Most importantly, damping of the induced oscillations is governed not only by $\gamma$ but also by the rate $D$, which is itself a function of the driving strength $\Omega$, leading to driving-dependent variations in the quality of oscillations as $\Omega$ is varied.

\begin{figure}[!t]
\centering
\includegraphics[width=3.0in,height=3.5in]{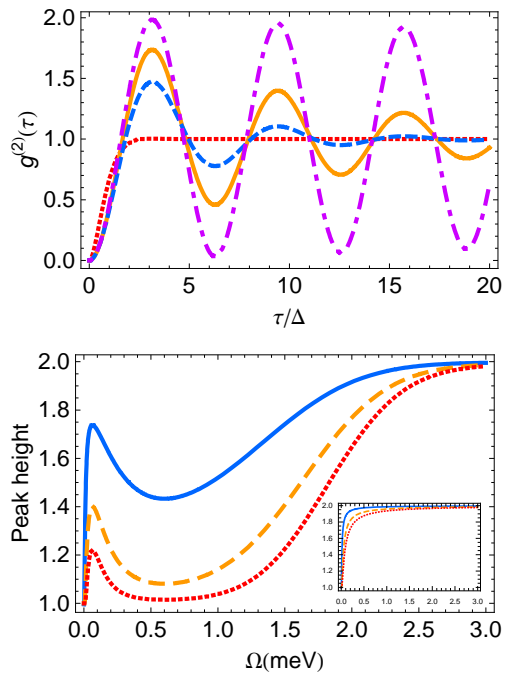}
\caption{(Color online) Top: $g^{(2)}$ against scaled delay time $\tau/\Delta$ for $\Omega=2$~$\mu{\rm eV}$ (red, dotted line), $\Omega=75$~$\mu{\rm eV}$ (yellow, solid line), $\Omega=0.4$~meV (blue, dashed line), and $\Omega=2.5$~meV (purple, dash-dot line). Bottom: Height of the first (blue, solid line), second (yellow, dashed line), and third (red, dotted line) peaks above as a function of $\Omega$. Inset: Height of corresponding peaks for a constant pure dephasing rate of $(1/150)$~$\rm{ps}^{-1}$. $\gamma=1/250$~${\rm ps}^{-1}$, all other parameters as in Fig.~\ref{figrates}.}
\label{figg2}
\end{figure}

This behaviour is illustrated in Fig.~\ref{figg2}~(a) where the correlation function is plotted as a function of delay time scaled by $1/\Delta$, to allow for easier comparison. We see that at the smallest driving frequency ($\Omega=2$~$\mu{\rm eV}$) no oscillations are yet visible and the function exhibits only the antibunching dip. As the driving strength is increased oscillations in $g^{(2)}(\tau)$ appear as a result of the induced coherent exciton evolution within the QD, shown here for $\Omega=75$~$\mu{\rm eV}$. These oscillations are damped due to dephasing of the excitonic state by phonons and as a result of spontaneous emission, though at least three periods are clearly observable. However, as the driving strength is further increased the quality of oscillations declines, as seen for $\Omega=0.4$~meV. This is in marked contrast to the expectation for a constant phenomenological pure dephasing rate $\Gamma_2^*$,~\cite{Carmichael93,Muller07,Batalov07} where the quality should simply increase with $\Omega$ due to a gain in Rabi frequency relative to the dephasing rate. As stated above, the origin of this effect is in the driving dependence of the dephasing rate $D$, itself a result of the form of the exciton-phonon interaction and its frequency dependence as seen in $J(\omega)$. For $\Omega=0.4$~meV, $D$ has grown sufficiently in comparison to $\Omega$ that the oscillations are now strongly damped. Only when $\Omega$ is increased to higher values, beyond the cut-off, does the damping weaken; in the case of $\Omega=2.5$~meV, $D$ is now small enough that the dephasing rate is dominated by $\gamma$. This point is emphasised in Fig.~\ref{figg2}~(b) where the heights of the three peaks in Fig.~\ref{figg2}~(a) are plotted as a function of $\Omega$. After an initial increase a dip is observed in the height of all peaks, and almost complete suppression in the case of the third. The width of this dip corresponds approximately to the width of the peak in $D$ (see Fig.~\ref{figrates}). Of course, as $\Omega$ increases further, the peak heights rise once more reaching their maximum value around $\Omega=3$~meV. Again, this is contrary to the behaviour expected for a constant pure dephasing rate, shown in the inset for $\Gamma_2^*=(1/150)$~$\rm{ps}^{-1}$, which exhibits no dip whatsoever.

\begin{figure}[!t]
\centering
\includegraphics[width=3.0in,height=3.5in]{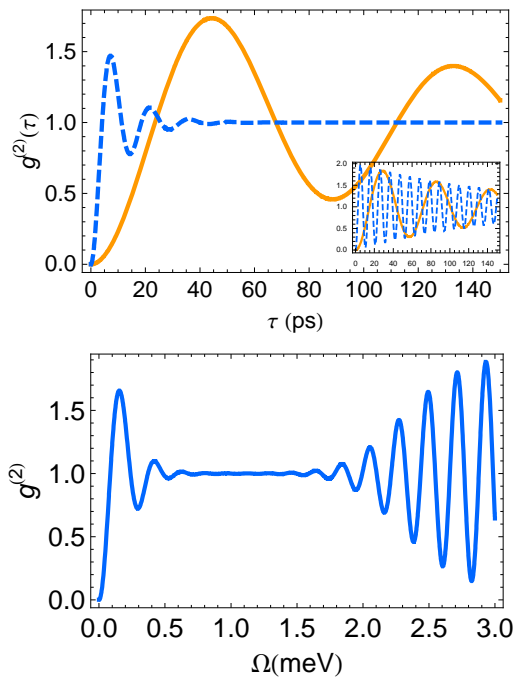}
\caption{(Color online) Top: $g^{(2)}$ against delay time $\tau$ for $\Omega=75$~$\mu{\rm eV}$ (yellow, solid line) and $\Omega=0.4$~meV (blue, dashed line). Inset: As main but for constant pure dephasing rate of $(1/150)$~$\rm{ps}^{-1}$. Bottom: Oscillations in $g^{(2)}$ as a function of $\Omega$ at fixed delay $\tau=20$~ps, akin to the Rabi rotations behaviour in Ref.~\onlinecite{Vagov07}. All other parameters are as in Fig.~\ref{figrates}.}
\label{figg2pt2}
\end{figure}

We can see then that for a significant range of the parameters considered, oscillations in $g^{(2)}(\tau)$ completely disappear after the first few picoseconds, resulting in independent photodetection events after this time delay. This effect is shown in Fig.~\ref{figg2pt2}~(a) where, for $\Omega=0.4$~meV, independent detection events are expected at delays beyond $\tau\sim40$~ps, while oscillations persist over a much longer timescale for weaker driving of $\Omega=75$~$\mu{\rm eV}$. As before, this can be contrasted to the case of a constant dephasing rate, illustrated in the inset. For both field strengths oscillations persist well beyond $100$~ps, decaying at the same rate. Finally, in Fig~\ref{figg2pt2}~(b) we see that by fixing the time delay, here at $20$~ps though the actual value is not crucial, the appearance, disappearance, and subsequent reappearance of oscillations in the correlation function can be observed as $\Omega$ is increased. This corresponds to a crossover from correlated to uncorrelated photodetection and back again; a striking effect with its origin solely in the driving dependence of the dephasing rate. Besides being of fundamental interest, this signature provides a means to map out the effective dephasing rate of the QD system, thus allowing for the determination of optimum performance conditions for the dot operating either as a qubit or single photon source. 

To summarise, direct measurements of the second-order fluorescence intensity correlation function of a coherently driven QD can be extremely sensitive to the internal dot coherence properties, going beyond what would be expected from more simplified models. I have shown that natural variations in the phonon-induced dephasing rate result in regimes of correlated and uncorrelated photon emission, dependent upon the driving field strength. This response should be observable in QDs by exploiting recently developed semiconductor and laser technology,~\cite{Krenner2005,Shields2007,Muller07} and could also be relevant to other solid-state systems.~\cite{Batalov07}

I would like to thank B. Lovett, T. Stace, S. Barrett, A. Kolli, E. Gauger, G. Pryde, D. Pegg, and H. Wiseman for engaging and informative 
discussions. I am supported by Griffith University, the State of Queensland, the Australian Research Council, and the {\sc EPSRC}.

\end{document}